\documentclass[a4paper]{article}
\usepackage[ansinew]{inputenc}
\usepackage{times}
\usepackage[T1]{fontenc}
\usepackage{graphicx}
\usepackage{geometry}
\usepackage{amssymb}
\usepackage{amsmath}

\usepackage{rotating}

\usepackage{xcolor}
\colorlet{shadecolor}{gray!25}
\usepackage{subfig}
\usepackage{float}
\usepackage{lscape}

\title{Simultaneous comparisons of treatments versus control (Dunnett-type tests) for location-scale alternatives}
\author{Ludwig A. Hothorn\\ (retired from) Leibniz University Hannover, Germany ($ludwig@hothorn.de$)}
				
\date{\today}

\begin{document}
\maketitle

\begin{abstract}
Commonly, the comparisons of treatment groups versus a control is performed for location effects only where possible scale effects are considered as disturbing. Sometimes scale effects are also relevant, as a kind of early indicator for changes. Here several approaches for Dunnett-type tests for location or scale effects are proposed and compared by a simulation study. Two real data examples are analysed accordingly and the related R-code is available in the Appendix.
\end{abstract}

\section{The problem}
The Dunnett test based on the common variance estimator and its degree of freedom of an one-way layout, i.e. the assumption of variance homogeneity holds. In practical data, heteroscedasticity occur sometimes which is considered as a disturbing influence factor. To avoid a violation of the FWER, modifications are recommended where the sandwich variance estimator is used \cite{Herberich2010}, or the degree of freedom is reduced accordingly \cite{Hasler2008}. In addition, there is a possible bias if, for example, the increased variance in a group of no interest which may biases the inference of another group of interest. Here we consider heterogeneity not primarily as a confounding factor, but a kind of early detection criteria before location effects become visible. This phenomenon can be seen, for example, in genetics, global warming, or income differences in the case of external stress, such as a pandemic. Here it is not about the detection of pure location effects on the one hand or pure scale effects on the other hand, but always about the detection of combined location and scale effects. 
\\
The basic approach is a max(maxT) test, simultaneously modeling both location effects and scale effects for simultaneous comparisons to the control.  For this purpose, an asymptotic multiple marginal models method (mmm) \cite{Pipper2012} is used. A scale-sensitive modified 2nd endpoint is defined by the Levene transformation $abs(y_{ij}-median(y_i))$ \cite{Neuhauser2000}. A further approach is resampling test using the Lepage test idea \cite{Hothorn2006, Hothorn2008b}. As an alternative, the most likely transformation method is used \cite{Hothorn2018, Hothorn2020a}, which accounts for location and scale differences in the context of more general distributional differences, again for a Dunnett-type test \cite{Hothorn2019}. For comparison (although considerably limited), Dunnett-type tests for location 
(and homogeneous variance), for location effects with sandwich estimators, and for scale (Levene residuals) are included.

\section{Location-Scale Dunnett-type tests}
First, max(maxT) test using the asymptotic mmm method to achieve the variance-covariance matrix of several model fits where $\xi=k*\eta$ models with a factor $A$ are included: $k$ for comparing of $k$ treatments versus control and $\eta=2$ for the untransformed endpoint $Y$ and the Levene-transformed endpoint $Yt$ (raw data available in the object $dat$) (MMM). The R-code shows this simple approach clearly:
\footnotesize
\begin{verbatim}
library(multcomp)
mod1<-lm(Y~A, data=dat)
mod2<-lm(tY~A, data=dat)
summary(glht(mmm(location = mod1, scale= mod2), mlf(mcp(A ="Dunnett"))))
\end{verbatim}
\normalsize  
Second, the most likely transformation approach (MLT) is more complex. A transformation using Bernstein polynomials is needed (where the order 5 is appropriate for such data). A conditional transformation model (ctm) is used for an additive factor shift and the transformation to normal distribution, followed by the most likely transformation (mlt). Finally, the parameter estimates are used in the generalize hypotheses testing function(glht) with the Dunnett-type correlation matrix.

\footnotesize
\begin{verbatim}
library(mlt)
yvar <- numeric_var("Number", support =quantile(dat$Y, prob = c(.01, .99))) # MLT support
bstorder<-5 # order of Bernstein polynomial (an appropriate choice)
yb <- Bernstein_basis(yvar, ui = "increasing",order =bstorder) # Bernstein polynomial
ma <- ctm(yb, shifting = ~ A, todistr = "Normal", data = dat) # condit transformation model
m_mlt<-mlt(ma, data = dat) # most likely transformation
K <- diag(length(coef(m_mlt))) # contrast matrix
rownames(K) <- names(coef(m_mlt)); matr<-bstorder+1
K <- K[-(1:matr),] # contrast matrix when using order 5 Bernstein
summary(glht(m_mlt, linfct = K)) # MLT-Dunnett-type test
\end{verbatim}
\normalsize

Third, the Lepage test approach (LEPA) can be modeled by the transformation function $legape_trafo$ in the library(coin) where the object DUN contains the Dunnett-type contrast matrix: \cite{Hothorn2006, Hothorn2008}

\footnotesize
\begin{verbatim}
library("coin")
lepage_trafo <- function(y)   cbind("Location" = rank_trafo(y), "Scale" = ansari_trafo(y))
pvalue(independence_test(Y ~ A, data = dat, xtrafo =DUN, teststat = "maximum",
                               distribution = approximate(nresample = 10000),
                               ytrafo = function(data)
                                 trafo(data, numeric_trafo = lepage_trafo)), method = "single-step")
\end{verbatim}

\normalsize
\section{Simulation results}
Both FWER (under global $H_0$) and the any-pairs power (under $H_1$) for the three basic location-scale Dunnett-type tests: MMM, MLT, LEPA
are reported in Table \ref{tab:sim} for a balanced design and an unbalanced design ($n_i=16,8,8,8$). Additional the estimates for the location models within mmm (MMMl), the standard Dunnett Test, the Dunnett-test modified with sandwich estimator (sDUN), the scale test (SCA) and the location models with Lepage tests (LEPAl) are considered. \textit{($H_{10}$ is for location effects only, $H_{01}$ is for scale effects only, $H_{11}$ is for location and scale effects, $H_{11}^d$ is for location and scale effects where the heterogeneity occurs in a non-interesting treatment group).} Variance heterogeneity was formulated by just increasing a treatment variance (MQR ... unadjusted) or keeping the MQR constant (MQR ... adjusted).\\
Per definition, FWER is controlled by all tests (MLT asymptotically only), however for any pattern of heteroscedasticity, the standard Dunnett test (DUN) behaves inflationary, where the sandwich modification still controls FWER. Per definition, MMM detects pure location effects, or pure scale effects or location and scale effects. MMM's power loss versus sDUN in the case of pure location effects, or versus SCA in the case of pure scale effects exists per definition, but is acceptable small. In most configurations, MMM outperforms both MLT and LEPA.

\begin{table}[ht]
\centering
\caption{Simulations under $H_{00}$ and $H_1$ for location-scale Dunnett-type tests}
\label{tab:sim}
\scalebox{0.6}{
\begin{tabular}{r|r|rrr|rrr||rrrrrrrr}
  \hline
MQR & Hypo. & $\mu_2$ & $\mu_3$ & $\mu_4$ & $s_2$ & $s_3$ & $s_4$ & MMM & MMMl & DUN & sDUN & SCA & MLT & LEPA & LEPAl \\ 
  \hline 
Unadj &	$H_{00}$ & 1.0 & 1.0 & 1.0 & 0.8 & 0.8 & 0.8 & 0.05 & 0.03 & 0.06 & 0.05 & 0.03 & 0.06 & 0.06 & 0.03 \\ 
\hline \hline
&$H_{01}$ & 1.0 & 1.0 & 1.0 & 0.8 & 0.8 & 1.8 & 0.39 & 0.04 & 0.06 & 0.05 & 0.46 & 0.04 & 0.23 & 0.02 \\ 
&$H_{01}$ & 1.0 & 1.0 & 1.0 & 0.8 & 0.8 & 2.4 & 0.66 & 0.06 & 0.10 & 0.05 & 0.73 & 0.07 & 0.40 & 0.04 \\ 
&$H_{01}^d$ & 1.0 & 1.0 & 1.0 & 1.8 & 0.8 & 0.8 & 0.40 & 0.03 & 0.05 & 0.05 & 0.46 & 0.04 & 0.23 & 0.03 \\ 
\hline
&$H_{10}$ & 1.0 & 1.0 & 2.5 & 0.8 & 0.8 & 0.8 & 0.93 & 0.93 & 0.96 & 0.94 & 0.04 & 0.96 & 0.84 & 0.82 \\ 
\hline
&$H_{11}$ & 1.0 & 1.0 & 2.5 & 0.8 & 0.8 & 1.8 & 0.74 & 0.58 & 0.66 & 0.47 & 0.47 & 0.54 & 0.48 & 0.32 \\ 
&$H_{11}$ & 1.0 & 1.0 & 2.5 & 0.8 & 0.8 & 2.4 & 0.82 & 0.43 & 0.52 & 0.31 & 0.72 & 0.37 & 0.54 & 0.23 \\ 
&$H_{11}^d$ & 1.0 & 1.0 & 2.5 & 1.8 & 0.8 & 0.8 & 0.80 & 0.66 & 0.79 & 0.94 & 0.44 & 0.85 & 0.79 & 0.72 \\ 
&$H_{11}^d$ & 1.0 & 1.0 & 2.5 & 2.4 & 0.8 & 0.8 & 0.81 & 0.41 & 0.54 & 0.93 & 0.72 & 0.73 & 0.83 & 0.67 \\ 
   \hline \hline
Adj & $H_{01}$ & 1.0 & 1.0 & 1.0 & 0.5 & 0.5 & 1.9 & 0.82 & 0.06 & 0.09 & 0.07 & 0.88 & 0.06 & 0.52 & 0.03 \\ 
&$H_{01}^d$ & 1.0 & 1.0 & 1.0 & 1.9 & 0.5 & 0.5 & 0.83 & 0.06 & 0.09 & 0.05 & 0.88 & 0.06 & 0.50 & 0.04 \\ 
&$H_{01}^d$ & 1.0 & 1.0 & 1.0 & 1.8 & 0.6 & 0.6 & 0.66 & 0.05 & 0.07 & 0.06 & 0.75 & 0.05 & 0.39 & 0.02 \\ 
&$H_{01}$ & 1.0 & 1.0 & 1.0 & 0.6 & 0.6 & 1.8 & 0.67 & 0.06 & 0.10 & 0.05 & 0.75 & 0.07 & 0.41 & 0.04 \\ 
   \hline
&$H_{11}$ & 1.0 & 1.0 & 2.5 & 0.5 & 0.5 & 1.9 & 0.96 & 0.65 & 0.72 & 0.49 & 0.89 & 0.53 & 0.73 & 0.35 \\ 
 & $H_{11}^d$& 1.0 & 1.0 & 2.5 & 1.9 & 0.5 & 0.5 & 0.98 & 0.76 & 0.86 & 1.00 & 0.88 & 0.98 & 0.99 & 0.96 \\ 
& $H_{11}^d$& 1.0 & 1.0 & 2.5 & 1.8 & 0.6 & 0.6 & 0.96 & 0.75 & 0.86 & 1.00 & 0.76 & 0.95 & 0.96 & 0.90 \\ 
& $H_{11}$& 1.0 & 1.0 & 2.5 & 0.6 & 0.6 & 1.8 & 0.89 & 0.64 & 0.73 & 0.49 & 0.74 & 0.55 & 0.62 & 0.35 \\ 
   \hline
	
Adj $n_0=16$ & $H_{10}$ & 1.0 & 1.0 & 2.5 & 0.8 & 0.8 & 0.8 & 0.94 & 0.94 & 0.96 & 0.95 & 0.04 & 0.96 & 0.72 & 0.69 \\ 
& $H_{11}$& 1.0 & 1.0 & 2.5 & 0.8 & 0.8 & 1.8 & 0.81 & 0.64 & 0.71 & 0.43 & 0.47 & 0.58 & 0.46 & 0.30 \\ 
&$H_{11}$ & 1.0 & 1.0 & 2.5 & 1.8 & 0.8 & 0.8 & 0.88 & 0.74 & 0.84 & 0.95 & 0.46 & 0.90 & 0.74 & 0.61 \\ 
   \hline	
	
\end{tabular}
}
\end{table}

\section{Evaluation of two examples using CRAN packages}

\subsection{Possible location and scale effects}
First an example was selected where both location and scale effects may occur. In a reverse genetic rescue experiment using Arabidopsis for the seed filling phenotype of acr4-2, the number of seeds per siliques in 10 transgenic lines were compared to wild type (wt), see the boxplots \cite{Okuda2020}. The related R-code is in Appendix I. The transgenic line f reveals the strongest decrease vs. wild type where both a locations effect and a scale effect was observed, see Table \ref{tab:gen}.

\begin{figure}[ht]
	\centering
		\includegraphics[width=0.3468\textwidth]{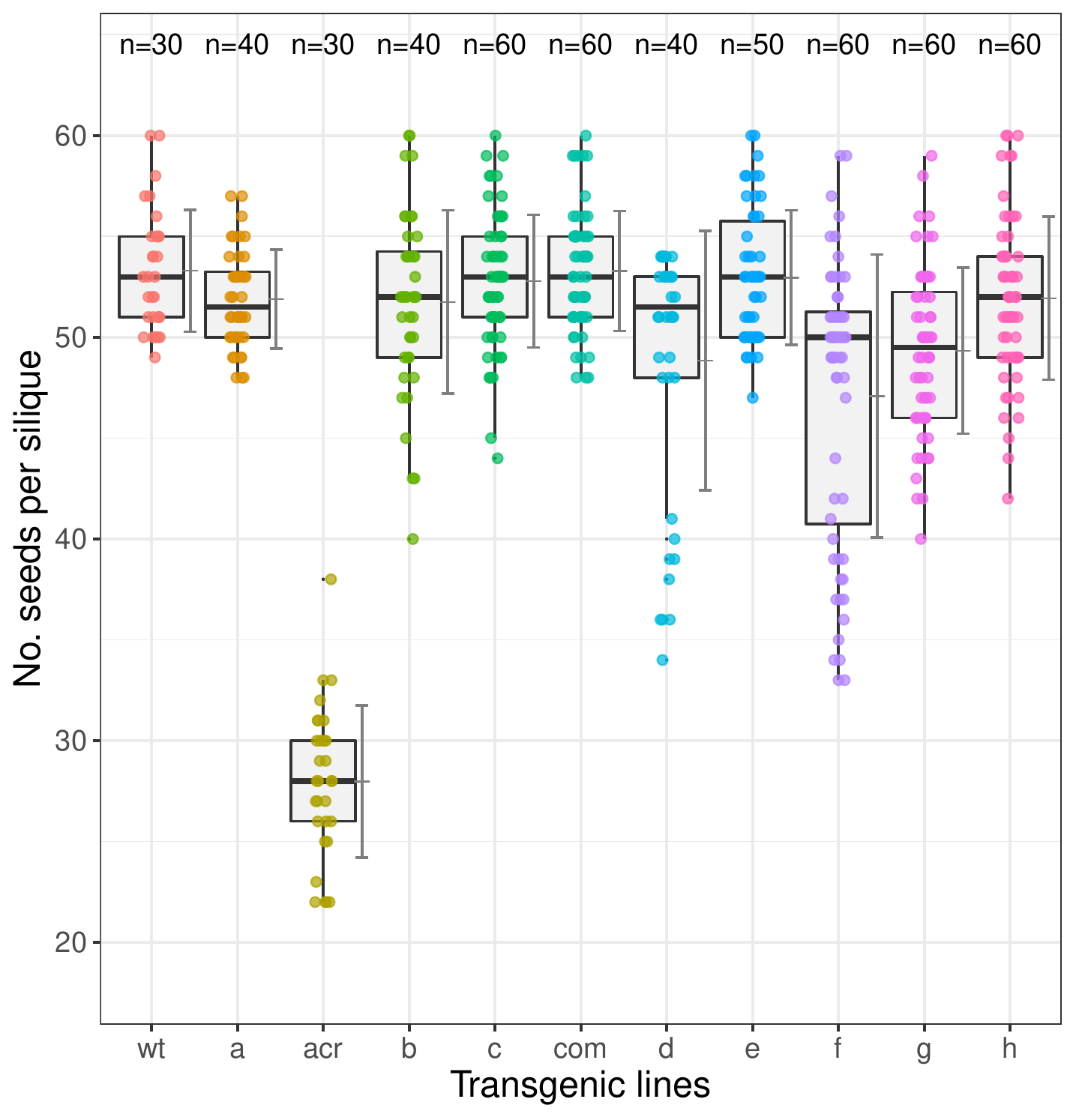}
	\caption{Number of seeds in a reverse genetic rescue experiment}
	\label{fig:F4}
\end{figure}

\begin{table}[ht]
\centering\scriptsize
\caption{Adjusted p-values of 4 approaches- reverse genetic rescue experiment}
\label{tab:gen}
\begin{tabular}{r|l|rr|rr}
  \hline
No & Comparison & Location only & Scale only & LocationScale & MLT \\ 
  \hline
1 & location: a - wt & 0.65 &-& 0.92 & 0.45\\ 
  2 & location: acr - wt & 0.0001 &-& 0.0001 & 0.0001 \\ 
  3 & location: b - wt & 0.55 &-& 0.85 & 0.59\\ 
  4 & location: c - wt & 0.99 &-& 0.99 & 0.99\\ 
  5 & location: com - wt & 0.99 &-& 0.99 &0.99\\ 
  6 & location: d - wt & 0.0001 &-& 0.0001 & 0.0013\\ 
  7 & location: e - wt & 0.99 &-& 0.99 &0.99\\ 
  8 & location: f - wt & 0.0001 &-& 0.0001 & 0.0001\\ 
  9 & location: g - wt & 0.0001 &-& 0.0001 & 0.0002\\ 
  10 & location: h - wt & 0.60 &-& 0.89 &0.56\\ \hline
  11 & scale: a - wt & - &0.99 &0.99 &-\\ 
  12 & scale: acr - wt & - &0.99 & 0.99 &-\\ 
  13 & scale: b - wt & - &0.66 & 0.92 &-\\ 
  14 & scale: c - wt & - &0.99 & 0.99 &-\\ 
  15 & scale: com - wt & - &0.99 & 0.99 &-\\ 
  16 & scale: d - wt & - &0.03 & 0.07 &-\\ 
  17 & scale: e - wt & - &0.99 & 0.99 &-\\ 
  18 & scale: f - wt & - &0.0001 & 0.0001 &-\\ 
  19 & scale: g - wt & - & 0.65& 0.92 &-\\ 
  20 & scale: h - wt & - &0.83 & 0.98 &-\\ 
   \hline
\end{tabular}
\end{table}

\subsection{Considering disturbing variance heterogeneity}
A second example was selected where a disturbing variance effect in a non-interesting dose group may occur. As example data, the cholesterol values in a chronic toxicity study was used \cite{Hothorn2016} where in the box-plots an increased variance in the low dose group occurred (global variance heterogeneity due to the Levene test in library(EnvStats). This low dose group is not of interest because only decreasing cholesterol values are of interest in this toxicological bioassay. The question arises which impact this low dose group has on the effect of the higher doses vs. control? The magnitude of global variance heterogeneity is not too large ($p_{Levene}=0.07$). In this example data, teh MLT approach detects an effect for the  250 mg dose group already.

\begin{figure}[ht]
	\centering
		\includegraphics[width=0.34\textwidth]{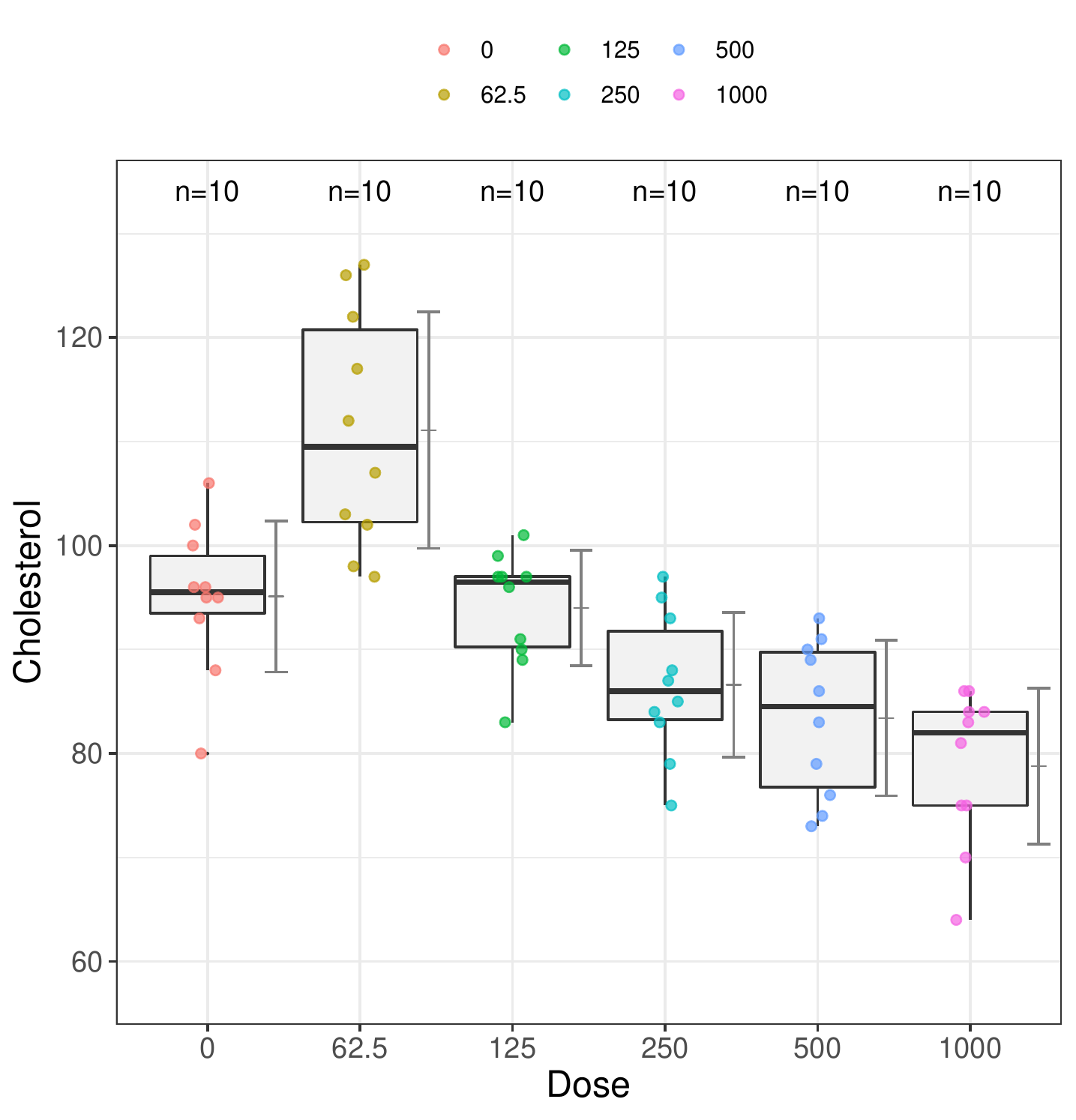}
	\caption{Cholesterol example data}
	\label{fig:Chol}
\end{figure}

\begin{table}[ht]
\centering\footnotesize
\caption{Adjusted p-values of 4 approaches- cholesterol experiment}
\begin{tabular}{rl|ll|l|l}
  \hline
No & Comparison & Location/Scale only & LocationScale& MLT & LocSandwich \\ 
  \hline
1 & location: 62.5 - 0 & 0.0001 & 0.0001&0.0006 &0.004\\ 
  2 & location: 125 - 0 & 0.99 & 0.99 &0.99& 0.99 \\ 
  3 & location: 250 - 0 & 0.08 & 0.12 &0.026& 0.058\\ 
  4 & location: 500 - 0 & 0.007 & 0.01 &0.0014& 0.0065\\ 
  5 & location: 1000 - 0 & 0.0001 & 0.0001 &0.0001& 0.0001\\ \hline
  6 & scale: 62.5 - 0 & 0.10 & 0.15 & &\\ 
  7 & scale: 125 - 0 & 0.99 & 0.99 & &\\ 
  8 & scale: 250 - 0 & 0.99 & 0.99 & &\\ 
  9 & scale: 500 - 0 & 0.99 & 0.99 & &\\ 
  10 & scale: 1000 - 0 & 0.99 & 0.99 & & \\ 
   \hline
\end{tabular}
\end{table}

\section{Conclusion}
If both location effects and scale effects might be of interest when comparing treatments to a control, a max(maxT) test using multiple marginal approach (mmm) is well suited. Here, the elementary decisions for both the individual treatment contrasts for both location and scale effects are available as adjusted p-values or simultaneous confidence intervals. The CRAN package multcomp makes the calculation relatively simple.

\scriptsize
\bibliographystyle{plain}

\newpage

\section{Appendix I: R-code example I}
\tiny
\begin{verbatim}
F4 <-
structure(list(Group = c("wt", "wt", "wt", "wt", "wt", "wt", 
"wt", "wt", "wt", "wt", "wt", "wt", "wt", "wt", "wt", "wt", "wt", 
"wt", "wt", "wt", "wt", "wt", "wt", "wt", "wt", "wt", "wt", "wt", 
"wt", "wt", "acr", "acr", "acr", "acr", "acr", "acr", "acr", 
"acr", "acr", "acr", "acr", "acr", "acr", "acr", "acr", "acr", 
"acr", "acr", "acr", "acr", "acr", "acr", "acr", "acr", "acr", 
"acr", "acr", "acr", "acr", "acr", "com", "com", "com", "com", 
"com", "com", "com", "com", "com", "com", "com", "com", "com", 
"com", "com", "com", "com", "com", "com", "com", "com", "com", 
"com", "com", "com", "com", "com", "com", "com", "com", "com", 
"com", "com", "com", "com", "com", "com", "com", "com", "com", 
"com", "com", "com", "com", "com", "com", "com", "com", "com", 
"com", "com", "com", "com", "com", "com", "com", "com", "com", 
"com", "com", "a", "a", "a", "a", "a", "a", "a", "a", "a", "a", 
"a", "a", "a", "a", "a", "a", "a", "a", "a", "a", "a", "a", "a", 
"a", "a", "a", "a", "a", "a", "a", "a", "a", "a", "a", "a", "a", 
"a", "a", "a", "a", "b", "b", "b", "b", "b", "b", "b", "b", "b", 
"b", "b", "b", "b", "b", "b", "b", "b", "b", "b", "b", "b", "b", 
"b", "b", "b", "b", "b", "b", "b", "b", "b", "b", "b", "b", "b", 
"b", "b", "b", "b", "b", "c", "c", "c", "c", "c", "c", "c", "c", 
"c", "c", "c", "c", "c", "c", "c", "c", "c", "c", "c", "c", "c", 
"c", "c", "c", "c", "c", "c", "c", "c", "c", "c", "c", "c", "c", 
"c", "c", "c", "c", "c", "c", "c", "c", "c", "c", "c", "c", "c", 
"c", "c", "c", "c", "c", "c", "c", "c", "c", "c", "c", "c", "c", 
"d", "d", "d", "d", "d", "d", "d", "d", "d", "d", "d", "d", "d", 
"d", "d", "d", "d", "d", "d", "d", "d", "d", "d", "d", "d", "d", 
"d", "d", "d", "d", "d", "d", "d", "d", "d", "d", "d", "d", "d", 
"d", "e", "e", "e", "e", "e", "e", "e", "e", "e", "e", "e", "e", 
"e", "e", "e", "e", "e", "e", "e", "e", "e", "e", "e", "e", "e", 
"e", "e", "e", "e", "e", "e", "e", "e", "e", "e", "e", "e", "e", 
"e", "e", "e", "e", "e", "e", "e", "e", "e", "e", "e", "e", "f", 
"f", "f", "f", "f", "f", "f", "f", "f", "f", "f", "f", "f", "f", 
"f", "f", "f", "f", "f", "f", "f", "f", "f", "f", "f", "f", "f", 
"f", "f", "f", "f", "f", "f", "f", "f", "f", "f", "f", "f", "f", 
"f", "f", "f", "f", "f", "f", "f", "f", "f", "f", "f", "f", "f", 
"f", "f", "f", "f", "f", "f", "f", "g", "g", "g", "g", "g", "g", 
"g", "g", "g", "g", "g", "g", "g", "g", "g", "g", "g", "g", "g", 
"g", "g", "g", "g", "g", "g", "g", "g", "g", "g", "g", "g", "g", 
"g", "g", "g", "g", "g", "g", "g", "g", "g", "g", "g", "g", "g", 
"g", "g", "g", "g", "g", "g", "g", "g", "g", "g", "g", "g", "g", 
"g", "g", "h", "h", "h", "h", "h", "h", "h", "h", "h", "h", "h", 
"h", "h", "h", "h", "h", "h", "h", "h", "h", "h", "h", "h", "h", 
"h", "h", "h", "h", "h", "h", "h", "h", "h", "h", "h", "h", "h", 
"h", "h", "h", "h", "h", "h", "h", "h", "h", "h", "h", "h", "h", 
"h", "h", "h", "h", "h", "h", "h", "h", "h", "h"), Number = c(57L, 
53L, 52L, 55L, 57L, 52L, 60L, 54L, 60L, 56L, 55L, 55L, 53L, 51L, 
52L, 51L, 55L, 51L, 54L, 54L, 49L, 53L, 50L, 51L, 50L, 50L, 51L, 
50L, 50L, 58L, 30L, 22L, 25L, 28L, 26L, 25L, 22L, 29L, 30L, 26L, 
27L, 29L, 26L, 32L, 31L, 27L, 28L, 28L, 23L, 33L, 30L, 33L, 30L, 
27L, 31L, 22L, 38L, 28L, 31L, 22L, 53L, 55L, 59L, 56L, 54L, 55L, 
56L, 52L, 52L, 54L, 49L, 52L, 57L, 51L, 51L, 51L, 51L, 49L, 49L, 
54L, 51L, 53L, 54L, 56L, 50L, 52L, 53L, 59L, 59L, 55L, 48L, 51L, 
53L, 50L, 55L, 55L, 60L, 59L, 51L, 48L, 55L, 52L, 51L, 54L, 52L, 
55L, 55L, 56L, 59L, 54L, 52L, 51L, 54L, 51L, 50L, 55L, 54L, 55L, 
52L, 48L, 53L, 56L, 57L, 50L, 54L, 54L, 57L, 55L, 55L, 51L, 53L, 
48L, 49L, 52L, 53L, 51L, 51L, 51L, 49L, 49L, 49L, 50L, 51L, 53L, 
51L, 54L, 50L, 48L, 50L, 48L, 52L, 55L, 51L, 53L, 53L, 53L, 50L, 
52L, 55L, 50L, 56L, 49L, 49L, 55L, 54L, 52L, 51L, 52L, 54L, 51L, 
51L, 50L, 60L, 52L, 54L, 54L, 54L, 56L, 55L, 59L, 47L, 43L, 45L, 
50L, 40L, 48L, 43L, 47L, 48L, 59L, 56L, 50L, 60L, 52L, 52L, 53L, 
52L, 49L, 56L, 52L, 52L, 55L, 52L, 50L, 53L, 48L, 51L, 49L, 53L, 
53L, 51L, 53L, 54L, 45L, 51L, 52L, 55L, 51L, 53L, 58L, 50L, 52L, 
55L, 52L, 49L, 49L, 53L, 51L, 52L, 53L, 53L, 57L, 48L, 58L, 52L, 
44L, 55L, 58L, 59L, 54L, 56L, 56L, 53L, 48L, 51L, 60L, 53L, 56L, 
50L, 49L, 56L, 57L, 51L, 54L, 54L, 54L, 50L, 54L, 53L, 59L, 54L, 
54L, 53L, 53L, 53L, 51L, 54L, 52L, 53L, 53L, 39L, 36L, 38L, 48L, 
36L, 36L, 34L, 41L, 39L, 40L, 51L, 51L, 54L, 51L, 54L, 52L, 49L, 
54L, 54L, 53L, 49L, 48L, 53L, 51L, 53L, 54L, 51L, 48L, 53L, 54L, 
50L, 50L, 53L, 57L, 53L, 56L, 50L, 50L, 54L, 53L, 56L, 53L, 53L, 
54L, 60L, 57L, 60L, 51L, 52L, 58L, 49L, 50L, 51L, 50L, 53L, 58L, 
50L, 53L, 50L, 54L, 58L, 59L, 53L, 58L, 55L, 54L, 50L, 49L, 51L, 
56L, 57L, 50L, 52L, 50L, 49L, 47L, 51L, 52L, 49L, 50L, 37L, 34L, 
34L, 39L, 36L, 39L, 35L, 40L, 42L, 37L, 50L, 48L, 50L, 59L, 49L, 
51L, 59L, 52L, 50L, 53L, 51L, 47L, 50L, 51L, 51L, 48L, 53L, 56L, 
49L, 55L, 48L, 49L, 50L, 50L, 52L, 57L, 54L, 53L, 49L, 53L, 55L, 
53L, 50L, 51L, 51L, 52L, 50L, 50L, 49L, 50L, 38L, 41L, 49L, 44L, 
33L, 42L, 37L, 38L, 39L, 33L, 49L, 53L, 55L, 50L, 52L, 48L, 48L, 
53L, 53L, 56L, 50L, 50L, 53L, 48L, 48L, 46L, 44L, 51L, 50L, 47L, 
46L, 42L, 44L, 42L, 50L, 51L, 49L, 46L, 47L, 46L, 46L, 50L, 49L, 
44L, 43L, 44L, 46L, 45L, 50L, 40L, 59L, 55L, 53L, 52L, 53L, 53L, 
53L, 56L, 58L, 55L, 45L, 51L, 49L, 46L, 47L, 47L, 52L, 52L, 51L, 
49L, 53L, 54L, 56L, 60L, 54L, 57L, 51L, 49L, 59L, 60L, 54L, 51L, 
54L, 49L, 54L, 52L, 53L, 52L, 53L, 59L, 54L, 53L, 51L, 50L, 56L, 
56L, 55L, 60L, 52L, 54L, 48L, 51L, 47L, 48L, 46L, 49L, 49L, 49L, 
55L, 51L, 52L, 56L, 51L, 50L, 53L, 52L, 53L, 59L, 49L, 51L, 45L, 
49L, 53L, 47L, 42L, 46L, 44L, 50L, 47L, 49L), group = structure(c(1L, 
1L, 1L, 1L, 1L, 1L, 1L, 1L, 1L, 1L, 1L, 1L, 1L, 1L, 1L, 1L, 1L, 
1L, 1L, 1L, 1L, 1L, 1L, 1L, 1L, 1L, 1L, 1L, 1L, 1L, 3L, 3L, 3L, 
3L, 3L, 3L, 3L, 3L, 3L, 3L, 3L, 3L, 3L, 3L, 3L, 3L, 3L, 3L, 3L, 
3L, 3L, 3L, 3L, 3L, 3L, 3L, 3L, 3L, 3L, 3L, 6L, 6L, 6L, 6L, 6L, 
6L, 6L, 6L, 6L, 6L, 6L, 6L, 6L, 6L, 6L, 6L, 6L, 6L, 6L, 6L, 6L, 
6L, 6L, 6L, 6L, 6L, 6L, 6L, 6L, 6L, 6L, 6L, 6L, 6L, 6L, 6L, 6L, 
6L, 6L, 6L, 6L, 6L, 6L, 6L, 6L, 6L, 6L, 6L, 6L, 6L, 6L, 6L, 6L, 
6L, 6L, 6L, 6L, 6L, 6L, 6L, 2L, 2L, 2L, 2L, 2L, 2L, 2L, 2L, 2L, 
2L, 2L, 2L, 2L, 2L, 2L, 2L, 2L, 2L, 2L, 2L, 2L, 2L, 2L, 2L, 2L, 
2L, 2L, 2L, 2L, 2L, 2L, 2L, 2L, 2L, 2L, 2L, 2L, 2L, 2L, 2L, 4L, 
4L, 4L, 4L, 4L, 4L, 4L, 4L, 4L, 4L, 4L, 4L, 4L, 4L, 4L, 4L, 4L, 
4L, 4L, 4L, 4L, 4L, 4L, 4L, 4L, 4L, 4L, 4L, 4L, 4L, 4L, 4L, 4L, 
4L, 4L, 4L, 4L, 4L, 4L, 4L, 5L, 5L, 5L, 5L, 5L, 5L, 5L, 5L, 5L, 
5L, 5L, 5L, 5L, 5L, 5L, 5L, 5L, 5L, 5L, 5L, 5L, 5L, 5L, 5L, 5L, 
5L, 5L, 5L, 5L, 5L, 5L, 5L, 5L, 5L, 5L, 5L, 5L, 5L, 5L, 5L, 5L, 
5L, 5L, 5L, 5L, 5L, 5L, 5L, 5L, 5L, 5L, 5L, 5L, 5L, 5L, 5L, 5L, 
5L, 5L, 5L, 7L, 7L, 7L, 7L, 7L, 7L, 7L, 7L, 7L, 7L, 7L, 7L, 7L, 
7L, 7L, 7L, 7L, 7L, 7L, 7L, 7L, 7L, 7L, 7L, 7L, 7L, 7L, 7L, 7L, 
7L, 7L, 7L, 7L, 7L, 7L, 7L, 7L, 7L, 7L, 7L, 8L, 8L, 8L, 8L, 8L, 
8L, 8L, 8L, 8L, 8L, 8L, 8L, 8L, 8L, 8L, 8L, 8L, 8L, 8L, 8L, 8L, 
8L, 8L, 8L, 8L, 8L, 8L, 8L, 8L, 8L, 8L, 8L, 8L, 8L, 8L, 8L, 8L, 
8L, 8L, 8L, 8L, 8L, 8L, 8L, 8L, 8L, 8L, 8L, 8L, 8L, 9L, 9L, 9L, 
9L, 9L, 9L, 9L, 9L, 9L, 9L, 9L, 9L, 9L, 9L, 9L, 9L, 9L, 9L, 9L, 
9L, 9L, 9L, 9L, 9L, 9L, 9L, 9L, 9L, 9L, 9L, 9L, 9L, 9L, 9L, 9L, 
9L, 9L, 9L, 9L, 9L, 9L, 9L, 9L, 9L, 9L, 9L, 9L, 9L, 9L, 9L, 9L, 
9L, 9L, 9L, 9L, 9L, 9L, 9L, 9L, 9L, 10L, 10L, 10L, 10L, 10L, 
10L, 10L, 10L, 10L, 10L, 10L, 10L, 10L, 10L, 10L, 10L, 10L, 10L, 
10L, 10L, 10L, 10L, 10L, 10L, 10L, 10L, 10L, 10L, 10L, 10L, 10L, 
10L, 10L, 10L, 10L, 10L, 10L, 10L, 10L, 10L, 10L, 10L, 10L, 10L, 
10L, 10L, 10L, 10L, 10L, 10L, 10L, 10L, 10L, 10L, 10L, 10L, 10L, 
10L, 10L, 10L, 11L, 11L, 11L, 11L, 11L, 11L, 11L, 11L, 11L, 11L, 
11L, 11L, 11L, 11L, 11L, 11L, 11L, 11L, 11L, 11L, 11L, 11L, 11L, 
11L, 11L, 11L, 11L, 11L, 11L, 11L, 11L, 11L, 11L, 11L, 11L, 11L, 
11L, 11L, 11L, 11L, 11L, 11L, 11L, 11L, 11L, 11L, 11L, 11L, 11L, 
11L, 11L, 11L, 11L, 11L, 11L, 11L, 11L, 11L, 11L, 11L), .Label = c("wt", 
"a", "acr", "b", "c", "com", "d", "e", "f", "g", "h"), class = "factor")), row.names = c(NA, 
530L), class = "data.frame")

######### Levene transformation 
medrs<-tapply(F4$Number, F4$group, median) # group-specific medians
medrsV<-medrs[as.integer(F4$group)] # median for each subject
F4$tNumber<- abs(F4$Number-medrsV) # Levene transformed endpoint
############ per location, per scale and mmm
library(multcomp); library(sandwich); library(ggplot2)
mod1<-lm(Number~group, data=F4)
mod2<-lm(tNumber~group, data=F4)
Joint1 <- glht(mmm(location = mod1, scale= mod2), mlf(mcp(group ="Dunnett")))
JT1<-summary(Joint1)
Loc <- glht(mod1, linfct=mcp(group ="Dunnett"))
L1<-summary(Loc)$test$pvalue
Sca <- glht(mod2, linfct=mcp(group ="Dunnett"))
S1<-summary(Sca)
###################### mlt
library(mlt)
yvar <- numeric_var("Number", support =quantile(F4$Number, prob = c(.01, .99))) # MLT
bstorder<-5 # order of Bernstein polynomial
yb <- Bernstein_basis(yvar, ui = "increasing",
                      order =bstorder) # Bernstein polynomial
ma <- ctm(yb, shifting = ~ group, todistr = "Normal", data = F4) # condit transf mod
m_mlt<-mlt(ma, data = F4) # most likely transformation
K <- diag(length(coef(m_mlt))) # contrast matrix
rownames(K) <- names(coef(m_mlt))
matr<-bstorder+1
K <- K[-(1:matr),] # for order 5 Bernstein
C<-glht(m_mlt, linfct = K) # MLT-Dunnett-type test
fortify(summary(C))
\end{verbatim}

\section{Appendix II: R-code example II}
\tiny
\begin{verbatim}
CHOL <-
  structure(list(Dose = c(0, 0, 0, 0, 0, 0, 0, 0, 0, 0, 62.5, 62.5,
                          62.5, 62.5, 62.5, 62.5, 62.5, 62.5, 62.5, 62.5, 125, 125, 125,
                          125, 125, 125, 125, 125, 125, 125, 250, 250, 250, 250, 250, 250,
                          250, 250, 250, 250, 500, 500, 500, 500, 500, 500, 500, 500, 500,
                          500, 1000, 1000, 1000, 1000, 1000, 1000, 1000, 1000, 1000, 1000
  ), Cholesterol = c(102, 100, 106, 95, 96, 80, 96, 88, 95, 93,
                     126, 127, 103, 112, 122, 107, 117, 102, 98, 97, 97, 101, 99,
                     83, 89, 97, 97, 96, 91, 90, 87, 93, 75, 84, 97, 83, 88, 79, 85,
                     95, 89, 93, 86, 83, 76, 79, 73, 91, 74, 90, 83, 86, 84, 84, 81,
                     64, 75, 70, 75, 86)), class = "data.frame", row.names = c(NA,
                                                                               60L))
CHOL$dose<-as.factor(CHOL$Dose)

library(EnvStats)
varGroupTest(Cholesterol ~ dose, data = CHOL, test = "Levene")
######### Levene transformation abs()
medrs<-tapply(CHOL$Cholesterol, CHOL$dose, median) # group-specific medians
medrsV<-medrs[as.integer(CHOL$dose)] # median for each subject
CHOL$tC<- abs(CHOL$Cholesterol-medrsV) # Levene transformation
############ mmm
mod1<-lm(Cholesterol~dose, data=CHOL)
DF<-anova(mod1)$Df[2]
mod2<-lm(tC~dose, data=CHOL)

Joint1 <- glht(mmm(location = mod1, scale= mod2), mlf(mcp(dose ="Dunnett")))
JT1<-summary(Joint1)
jjt<-fortify(JT1)[, c(1,5,6)]
##### location, scale separate
Loc <- glht(mod1, linfct=mcp(dose ="Dunnett"), df=DF)
L1<-summary(Loc)
Loc2 <- glht(mod1, linfct=mcp(dose ="Dunnett"), vcov=vcovHC, df=DF)
L2<-summary(Loc2)
Sca <- glht(mod2, linfct=mcp(dose ="Dunnett"))
S1<-summary(Sca)
############### mlt to normal
library(mlt); library(ggplot2)
yvar <- numeric_var("Cholesterol", support =quantile(CHOL$Cholesterol, prob = c(.01, .99))) # MLT
bstorder<-5 # order of Bernstein polynomial
yb <- Bernstein_basis(yvar, ui = "increasing",
                      order =bstorder) # Bernstein polynomial
ma <- ctm(yb, shifting = ~ dose, todistr = "Normal", data = CHOL) # condit transf mod
m_mlt<-mlt(ma, data = CHOL) # most likely transformation
K <- diag(length(coef(m_mlt))) # contrast matrix
rownames(K) <- names(coef(m_mlt))
matr<-bstorder+1
K <- K[-(1:matr),] # for order 5 Bernstein
C<-glht(m_mlt, linfct = K) # MLT-Dunnett-type test
fortify(summary(C))
\end{verbatim}

\end{document}